\documentclass{article}
\usepackage{amsmath,amssymb,amsfonts}
\usepackage[dvips]{epsfig}

\newtheorem{theorem}{Theorem}[section]

\newtheorem{definition}[theorem]{Definition}
\newtheorem{corollary}[theorem]{Corollary}
\newtheorem{axiom}[theorem]{Axiom}
\newtheorem{proposition}[theorem]{Proposition}

\font\SYM=msbm10 
\newcommand{\Real}{\mbox{{\SYM R}}}

\newcommand{\dd}{\mbox{d}}

\font\tenscr=rsfs10 scaled1100
\font\sevenscr=rsfs7 
\font\fivescr=rsfs5 
\skewchar\tenscr='177 
\skewchar\sevenscr='177
\skewchar\fivescr='177 
\newfam\scrfam 
\textfont\scrfam=\tenscr
\scriptfont\scrfam=\sevenscr 
\scriptscriptfont\scrfam=\fivescr

\def\O{\mathcal{O}}
\def\Lie{\mathcal{L}}

\def\rhot{\widetilde{\rho}}

\def\ttilde{\widetilde{t}}
\def\zt{\widetilde{z}}

\def\bt{\widetilde{b}}

\def\scrd{{\fam\scrfam D}}

\title{The Newtonian limit of spacetimes describing uniformly accelerated
 particles}
\author{ Ruth Lazkoz \footnote{email address: {\tt wtplasar@lg.ehu.es}}\\
\emph{\small Fisika Teorikoaren eta Zientziaren Historiaren Saila } \\
\emph{\small Euskal Herriko Unibertsitatea} \\
\emph{\small 644 Posta Kutxatila} \\
\emph{\small 48080 Bilbao, Spain} \\
Juan Antonio Valiente Kroon \footnote{email address: {\tt
      jav@aei-potsdam.mpg.de}} \footnote{Current address: Institut
      f\"{u}r Theoretische Physik der Universit\"{a}t Wien,
      Boltzmanngasse 5, A-1090, Austria} \\
\emph{\small Max-Planck Institut f\"{u}r Gravitationsphysik,} \\
\emph{\small Albert Einstein Institut,} \\
\emph{\small Am M\"{u}hlenberg 1,} \\
\emph{\small 14476 Golm, Germany.}
}

\begin{document}

\maketitle

\begin{abstract}
  We discuss the Newtonian limit of boost-rotation symmetric
  spacetimes by means of the Ehlers' frame theory. Conditions
  for the existence of such a limit are given and, in particular, we
  show that asymptotic flatness is an essential requirement. 
  Consequently, generalized boost-rotation
  symmetric spacetimes describing particles moving in uniform fields
  will not possess such a limit. In the cases where the boost-rotation
  symmetric spacetime is asymptotically flat and its Newtonian limit
  exists, then the (Newtonian) gravitational potential agrees with the
  potential suggested by the weak field approximation. We illustrate
  our discussion through some examples: the Curzon-Chazy particle
  solution, the generalized Bonnor-Swaminarayan solution, and the C
  metric.
\end{abstract}

\bigskip
PACS: 04.20.Jb, 04.25.Nx, 04.20.Cv
\bigskip
\section{Introduction}
Boost-rotation symmetric spacetimes can be thought of as describing uniformly
accelerated particles. The uniform acceleration can in some cases be
interpreted as due to an external field, and in other cases as the outcome of
self-accelerations produced by the presence of positive an negative
masses, or even as the effect of a strut connecting pairs of particles. 
Precisely these last two types of models 
comprise the only known classes of exact solutions to the
Einstein field equations which are locally asymptotically flat, in the sense
that they possess sections of null infinity which are spherical, but null
infinity is not complete because some of its generators are not complete.

Boost-rotation symmetric spacetimes possess two (hypersurface orthogonal commuting) Killing vectors.
One of them is an axial Killing vector. The other one leaves invariant the
light cone through the origin, and can be regarded as the curved spacetime
generalisation of the boost Killing vectors of Minkowski spacetime. The boost
symmetry has a special status, being the only other symmetry a radiative
axially symmetric spacetime can have \cite{BicSch84,BicPra98,Val00b}.

Historically, these spacetimes were of outmost importance for it 
was a solution of this kind \cite{BonSwa64} that became the first 
explicit non-stationary solution describing gravitational radiation 
according to Bondi's description  and Penrose's treatment of asymptotic
flatness ---\cite{Bic68}---.

A procedure to construct systematically boost-rotation symmetric spacetimes
 both in the case describing freely falling particles
\cite{BicHoeSch83a}, and in the case describing self-accelerated
particles \cite{BicHoeSch83b} has been given. Bi\v{c}\'{a}k \& Schmidt
\cite{BicSch89a} have provided a unified discussion of those boost-rotation
symmetric spacetimes which are as ``asymptotically flat as possible''. There,
it was shown that in order to obtain all the spacetimes in the
class, one has to start by solving an inhomogeneous wave equation in flat
space with sources moving along the orbits of the boost rotation
Killing vector. These solutions to the inhomogeneous wave equation
were then used as ``seeds'' for the boost-rotation space-times.

The boost-rotation symmetric spacetimes have been considered extensively in the
literature. For a discussion of their role in the understanding of the theory
of General Relativity see \cite{Bic00}. Some more specific studies can be
found in \cite{Bic85,Bic87,PraPra00}. However, the discussion of their
Newtonian limit has been only carried out from the weak field approximation
perspective, and so the question of the validity of the obtained results has
so far remained open in the lack of a more rigorous treatment.  It is the
purpose of the present article to address this very issue.

The study of the relationship between Newton's and Einstein's theories
of gravitation, which are respectively predecessor and successor
of one another, has lead to attempts to relate them
under certain limiting conditions. Despite the fact the  formulation 
of these two theories was grounded on very different concepts, the
existence of a common description setup allows one to recognize
Newton's theory as a degenerate limit of that of Einstein. 
By working in this common framework it would in principle be
possible to exploit some structural similarities and generalize
theoretical results from the old theory to the new one.

Perhaps the main motivation for carrying out this sort of investigations
is the fact that the  observational consequences of General Relativity
strongly rely on post-Newtonian approximations, or in other words, 
that the experimental
refutations of General Relativity are usually reported in the language
of Newton's theory. Summarizing, the predecessor theory
is an invaluable tool for devising approximation schemes allowing
to establish links between the two theories.

As discussed  above, one of the fundamental steps towards a 
well grounded link between the two theories is to give a precise definition
of the relevant approximations. Nowadays, the vinculum between the equations 
of General Relativity and Newton's theory is 
well understood. Yet the link between solutions to the equations
remains not completely clear. The difficulty lies in the difference
between the geometric notions used in the Newtonian setup and in the
General Relativity theory. Several schemes have been devised to write
both theories in a common language ---see for example \cite{Dau97,Win83,Ehl98}
and references therein--- so that the
transition from the relativistic theory to that of Newton can be
taken in a conceptually consistent way. Among them, Ehlers' theory is of particular interest due to its
covariant nature; this approach has been further developed by Lottermoser
---see for example \cite{Ehl98,Lot88} for full references---. We will  use
 this particular framework in our study of the Newtonian limit of the
boost-rotation symmetric spacetimes.

This article is structured as follows: in section 2 we begin by describing
briefly Ehlers' frame theory. In particular we focus on 
some technical results that will be used in our investigation of
the Newtonian limit. In section 3 we proceed to a general discussion of
boost-rotation symmetric spacetimes in a way which is suited for our later
discussion. In section 4 the discussion of the Newtonian limit of
boost-rotation symmetric spacetimes is actually carried out. A couple of
propositions regarding the conditions needed to have such limit are here
stated and proved. The role of asymptotic flatness in the existence of the
Newtonian limit is discussed. Finally, we address some interpretational issues,
in particular those of the determination of the proper Newtonian potential and
Newtonian sources.  In section 5 we analyse some examples: the Curzon-Chazy
particle solution, the generalized Bonnor-Swaminarayan, and the C metric.
Finally, an appendix containing an adaptation of the axioms of the frame
theory as given by J. Ehlers ---which are not so readily available in the
literature--- is included.

\section{Ehlers' frame theory}

First, we proceed to overview briefly the concepts, ideas and results of
Ehlers' frame theory in the form given by J. Ehlers ---see e.g.
\cite{Ehl98}--- that will be used in our investigation. For completeness, and
as a quick reference, an adapted version of the axioms of the frame theory is
given in the appendix to this article.  The required propositions and theorems
will be stated without proof. For a thorough discussion and the full details of
the proof we remit the reader   to \cite{Lot88}.

Ehlers' frame theory considers a 4-dimensional
differentiable manifold $M$ endowed with a torsion free connection 
(not necessarily metric) on which the two symmetric rank 2 tensors 
$t_{ij}$
(temporal metric) and $s^{ij}$ (spatial metric) are defined. The
temporal metric and the spatial metric are related to each other via
\begin{equation}
\label{inverses}
t_{ij}s^{jk}=-
\lambda\delta_{i}^{\phantom{i}k},
\end{equation}
where $\lambda \geq 0$ is a constant known as the \emph{causality
constant}. Throughout this work 
all Latin indices will range from $0$ to
$3$, 
except for $a$, that will range from $1$ to $3$. The summation
convention is assumed. If $\lambda \neq 0$, then the causality
constant can be identified with $1/c^2$, $c$ being the speed of light.
The temporal metric and the spatial metric are compatible with the
connection in the sense that
\begin{equation}
\nabla_k t_{ij}=0, \qquad \nabla_k s^{ij}=0.
\end{equation}

The spacetime manifold $M$ can be thought of as being parameterised by
the causality constant $\lambda$, so that in fact $M=M(\lambda)$
describes a family of spacetime manifolds. For $\lambda\neq 0$ a
change in the value of $\lambda$ can be regarded as a change of the
units in which the speed of light is measured. Intuitively, one would
wish to identify the Newtonian limit of the family $M(\lambda)$ with
$\lambda=0$. This limit will be of a degenerate nature as can be seen
from considering relation (\ref{inverses}). Following Ehlers, we make
the following definition:

\begin{definition} \label{def:newtonian_limit} \textbf{(Newtonian
    limit of a spacetime).} The family $M=M(\lambda)$ of spacetime
  manifolds is said to have a \emph{Newtonian limit} if:
\begin{itemize} 
\item[(i)] the connection, the spatial metric, the temporal metric, and the
Riemann tensor constructed from the  connection have a limit for
$\lambda\to 0$; 
\item[(ii)] the limiting value  of the connection,
  the Riemann tensor, the spatial metric and the temporal metric as $\lambda\to 0$ satisfy the
  axioms of the frame theory.
\end{itemize}
\end{definition}

One can easily understand that the spatial metric and the temporal metric 
are required to have a Newtonian limit by simply recalling that they are the 
fundamental objects of our theory. Similarly, the same must hold
for the connection, which in 
the limit $\lambda\to0$ should (intuitively) yield the gravitational 
potential. The equivalent requirement on the Riemann tensor is not so clear
though, the rationale behind it being that the Riemann tensor describes 
the effects  of  non-homogeneous
gravitational fields: the tidal forces. From this point of view, it is natural
to demand the Newtonian limits of spacetimes to have well defined tidal
forces.

For $\lambda\neq 0$, it is not difficult to relate the temporal metric and
the spatial metric to the metric tensor of General Relativity and its
inverse. For $\lambda=0$, this ceases being the case and the spacetime
acquires a degenerate metric structure. In order to study the
behaviour of the connection under these circumstances, it is convenient
to perform a 1+3 decomposition of it. This decomposition requires the
introduction of a timelike congruence $u^i$ (four-velocity
field of an observer) which on passing to the
Newtonian limit will give rise to the Galilean simultaneity surfaces. Note that
contrary to the case of (globally) stationary spacetimes, where a
canonical choice for such an observer field exists (the flow lines of
the timelike Killing vector), in radiative spacetimes such a canonical
choice does not  exist \emph{a priori}. 

The choice of a (normalized) observer field $u^i$ induces in
a natural way the projection tensor
$\pi^i_j=\delta^i_j-u^i\omega_j$, where we
have defined $\omega_i=u^\bullet_i=t_{ij}u^j$.
Here, and in what follows, the bullet ${}^\bullet$ indicates that the
corresponding tensorial object has been constructed by lowering 
indices using the temporal metric $t_{ij}$. One can also define
the tensors
\begin{eqnarray} 
&&\gamma_{ij}=\frac{1}{\lambda}(u^\bullet_i
u^\bullet_j-t_{ij}), \\
&&\kappa^{ij}=s^{ij}+\lambda u^i u^j.
\end{eqnarray}
A ``hydrodynamic''decomposition of the
derivative of the covariant derivative $u^i$ arises naturally:
\begin{equation}
\label{hydrodynamic}
u^i_{;j}=\kappa^{ik}\left(E_k\omega_j+\sigma_{kj}+
\frac{1}{3}\vartheta\gamma_{kj}+\Omega_{kj}\right),
\end{equation}
where $E_i$ is the 3-acceleration, $\sigma_{ij}$ the shear,
$\vartheta$ the expansion and $\Omega_{ij}$ the vorticity of the
observer field. It can be shown that if
$\Lie_u\gamma_{ij}=0$, where $\Lie_u$ denotes the Lie
derivative along $u^i$ then
$u^i_{;j}=\kappa^{ik}(E_k\omega_j+\Omega_{kj})$.
This type of observers will be called \emph{rigid}. The field
$u^i$ induces, as well, a certain decomposition on the
connection. In particular, one has the following result:

\begin{theorem}
\label{theorem:connection}
Let $u^i$ be a normalized observer (timelike) field. Let us define
\begin{equation}
F_{ij}=\frac{1}{\lambda}u^\bullet_{[j,i]}\mbox{
  (i.e. }F=\frac{1}{\lambda}\dd u^\bullet.)
\end{equation}
If $\lambda=0$, $\Lie_u \gamma_{ij}=0$, and the Newtonian limit of
$M(\lambda)$ exits then,

\begin{itemize}

\item[(i)] locally there exists a scalar function $\eta$ (absolute
  time) such that,
\begin{equation}
\omega_i=\eta_{,i};
\end{equation}

\item[(ii)] the connection can be written as,
\begin{eqnarray}
&&\Gamma^i_{jk}=\overline{\Gamma}^i_{jk}+
2\delta^{i a}F_{a(j}\eta_{,k)}\nonumber \\
&&\phantom{\Gamma^i_{jk}}=\overline{\Gamma}^i_{jk}+
\delta^{i a}\left( 2\Omega_{a(j}\eta_{,k)}+
   E_{a}\eta_{,j}\eta_{,k}\right)
\end{eqnarray}
where,
\begin{equation}
\overline{\Gamma}^i_{jk}=
\frac{1}{2}s^{il}\pi^m_j\pi^n_
k(\gamma_{l m,n}+\gamma_{n l, m}-
\gamma_{m n,\, l})
\end{equation}
and $E^i$ and $\Omega_{ij}$ are defined via the hydrodynamic
decomposition (\ref{hydrodynamic}).
\end{itemize}

\end{theorem}

Moreover, one has the following corollary, which indicates which parts of the
limit of the tensor $F_{ij}$ correspond to the (Newtonian)
gravitational potential ($E_a$), and which to a Coriolis field
($\Omega_{ij}$). This corollary will be our main interpretational tool. 

\begin{corollary}
\label{corollary:connection}
Choosing a Cartesian coordinate system, then (in the $\lambda=0$ case) the geodesics satisfy
\begin{equation}
\ddot{\vec{x}}=\vec{g}+2\dot{\vec{x}}\times\vec{\Omega},
\end{equation}
where
\begin{equation}
\vec{g}=-(E_1,E_2,E_3) \qquad \vec{\Omega}=
-(\Omega_{23},\Omega_{31},\Omega_{12}).
\end{equation}
\end{corollary}

There are several equivalent ways of verifying that a given family of 
spacetimes $M=M(\lambda)$ possess a Newtonian limit. The procedure that 
will be used in the present work is summarized in the following theorem.

\begin{theorem}
\label{existence:check}
Let $M(\lambda)$ be a family of spacetimes parameterised by $\lambda \geq 0$.
 Then if
\begin{itemize}
\item[(a)] \textbf{(Newtonian limit of the metrics)}
           \begin{itemize}
           \item[(a1)] the limit of $s^{ij}$ as $\lambda\to 0$
 exists and is of rank 3;
           \item[(a2)] $\lim_{\lambda\to 0}
\displaystyle{ \frac{1}{\lambda}}
 \det s^{ij}$ exists;
           \end{itemize}

\item[(b)]\textbf{(Newtonian limit of the connection)} given an observer field
 $u^i=u^i(\lambda)$ in $M(\lambda)$ such that 
$\lim_{\lambda\to 0} u^i$
 exists and
\begin{equation}
\lim_{\lambda\to 0} u^{i}_{,j}\,\exists =
(\lim_{\lambda\to 0}u^{i} )_{,j},
          \end{equation}        
one has
          \begin{itemize}
          \item[(b1)] 
          \begin{equation}
          \lim_{\lambda\to 0} t_{ij,k}\,\exists=
(\lim_{\lambda\to 0}t_{ij})_{,k}, \qquad 
 \lim_{\lambda\to 0} {s^{ij}}_{,k}\,\exists=
(\lim_{\lambda\to 0}s^{ij} )_{,k},
          \end{equation}
          \item[(b2)] 
          \begin{equation} 
           \lim_{\lambda\to 0}
           \frac{1}{\lambda}t_{[i|j |} {u^j}_{,k]}\,\exists =
           \lim_{\lambda\to 0} F_{ik};
          \end{equation}
          \end{itemize} 
\item[(c)] \textbf{(Newtonian limit of the Riemann tensor)} 

\begin{equation} 
           \lim_{\lambda\to 0}
           \frac{1}{\lambda}F_{ij,k}\,\exists =
           (\lim_{\lambda\to 0} F_{ij})_{\!,k};
          \end{equation}
\end{itemize}
then the family of spacetimes $M(\lambda)$ has a Newtonian limit.
\end{theorem}  

In this theorem the symbol $\exists =$ should be understood as meaning
``exists and is equal to''. A final notational remark: given a quantity
$x$ (scalar, tensor) such that its limit as $\lambda\to 0$
exists, we will often write $\overset{\circ}{x}$ for
$\lim_{\lambda\to 0}x$.

\section{Boost-rotation symmetric spacetimes}

Let us recall the customary interpretation of boost-rotation symmetric 
spacetimes as describing uniformly
accelerated particles. As is well known, in General Relativity the causes
of the motion are included in the theory. In our case, the acceleration can be
either due to an uniform external field or the effect of repulsion between
particles with positive and negative masses \footnote{For a discussion of the
  concept of negative mass in General Relativity, the reader is remitted to
  the classical article by Bondi \cite{Bon57}.}. Conical singularities
could also be the cause of the accelerations. Boost-rotation symmetric
spacetimes contain generally (naked) strut singularities. This feature makes
them in a way not very physical \footnote{See however \cite{HawRos95},
  in which it is argued that conical singularities arising in, for
  example, the C metric can be considered as a limit of real strings.}. However, those containing what can be
described as repelling pairs of positive
and negative masses constitute the only explicit  examples (in the realm of exact
solutions) of  locally asymptotically flat radiative spacetimes in the
sense of Penrose \cite{Pen63}. The particles in the spacetime undergo uniform
acceleration. Thus, they approach the speed of light asymptotically. Now, the
smoothness of the spacetime requires that it possesses reflection
symmetry. All this implies that null infinity has at least two singular
points. Therefore, some of the generators of null infinity are not complete.
This seems to be the best that can ever been achieved by means of
exact radiative solutions.

\begin{figure}[t]
\begin{center}
\label{figure:boost}
\includegraphics[width=.5\textwidth]{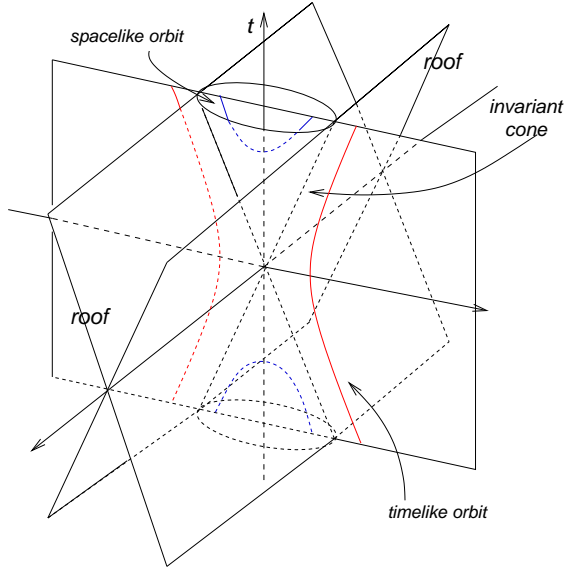}
\end{center}
\caption{The action of the boost Killing vector on flat spacetime
  ---see \cite{BicSch89a}---.}
\end{figure}

As we said, boost-rotation symmetric spacetimes have two commuting 
hypersurface orthogonal Killing
vector fields. One of them is an axial Killing vector ($\eta^i$),
and the other is the generalization of the boost Killing vector of
Minkowski spacetime ($\xi^i$), see figure 1. The
Killing vector $\xi^i$ leaves invariant the light cone through the
origin. As is the case in the Minkowski spacetime, there are
regions where the boost Killing vector is timelike ($\xi_i
\xi^i>0$), null ($\xi_i \xi^i=0$), and spacelike
($\xi_i \xi^i<0$). The set for which $\xi^i$ is null
will be known as the \emph{roof} ---see e.g. \cite{BicSch89a}---. Boost-rotation symmetric spacetimes
are time symmetric, thus we will only consider in our discussion the region
for which $t\geq 0$, as depicted in figure \ref{figure:boost}. The region of the
spacetime for which $t\geq 0$ and $\xi^i$ is timelike will be known
as \emph{below the roof}, whereas the portion for which $\xi^i$
is spacelike will be denoted as \emph{above the roof}. As will be
seen later, our discussion of the Newtonian limit of the
boost-rotation symmetric spacetimes will  naturally deal with  above the roof region
of the spacetime.
In the region below the roof, the boost-rotation symmetric spacetimes
can be locally put into the Weyl form, whereas above the roof, the spacetime
can be put locally in the form of an Einstein-Rosen wave. The
difference between the Weyl and Einstein-Rosen spacetimes and the
boost-rotation symmetric spacetimes arises when we consider their
global structure. Above the roof, the line element of a boost-rotation
symmetric spacetime can be written in the following form:
\begin{equation}
\label{d_line_element}
\dd s^2=\frac{1}{c ^2t^2-z ^2}\{(e^\nu z^2-e^\mu
c ^2t^2)\dd z^2+c ^2[2zt(e^\mu-e^\nu)\dd z \dd t+
(c ^2e^\nu t^2-e^\mu z^2)\dd t^2]\}
-e^\nu \dd\rho^2-\rho^2e^{-\mu}\dd\varphi^2,
\end{equation}
the functions $\mu$ and $\nu$ \footnote{In this article the function $\nu$
  will correspond to the function $\lambda$ of Bi\v{c}\'{a}k \& Schmidt
  \cite{BicSch89a} and most of the classical references on the subject. This
  is because $\lambda$ will be reserved for the causality constant. The reader
  has been warned!}
have the following functional dependence:
\begin{eqnarray}
&&\mu=\mu(\rho^2,t^2-z^2/c^2), \\
&&\nu=\nu(\rho^2,t^2-z^2/c^2).
\end{eqnarray}
In the $(t,\rho,\varphi,z)$ coordinates the region above the roof
corresponds to $c ^2 t^2>z^2$. If (\ref{d_line_element}) corresponds to 
a vacuum solution of Einstein equations then the function $\mu$ 
satisfies the wave equation
\begin{equation}
\label{wave}
\Box\mu=0,
\end{equation}
while $\nu$ can be found by quadratures once $\mu$ has been
obtained. 

Note now that coordinates $\rho$ and $z$ ($[\rho]=[z]=L$) from line
element (\ref{d_line_element}) have dimensions of length, while $t$ has
dimensions of time ($[t]=T$). In order to ease our discussion of the Newtonian
limit, it will be convenient to make use of dimensionless coordinates. To this
end, we assume that our system possesses a characteristic length $\alpha$
and a characteristic time $\tau$. The introduction of a characteristic time in
the relativistic regime of the frame theory is superfluous because from a given
length one can always construct a time just dividing by the speed of light.
However, in the Newtonian limit such canonical choice of speed does not longer
exist. This small redundancy is a price one has to pay in order to write the
two theories in a common language. Dimensionless coordinates are then given by
\begin{eqnarray}
&& \rhot=\alpha^{-1}\rho, \\
&& \zt=\alpha^{-1}z, \\
&& \ttilde=\tau^{-1}t,
\end{eqnarray} 
$\alpha$ and $\tau$ being constants such that $[\alpha]=L$ and $[\tau]=T$.

In the sequel, it will be convenient to use above the roof
coordinates $(\bt,\rhot,\varphi,\tilde\chi)$, which 
diagonalize the line
element of the spacetime \cite{BicSch89a}:
\begin{equation}
\label{nd_line_element}
\dd s^2=\frac{\tau^2}{\lambda}e^\nu \dd \bt^2-\alpha ^2
\left(e^\nu \dd\rhot^2-\rhot^2e^{-\mu}\dd\varphi^2-
{\bt^2}e^\mu \dd\tilde\chi^2\right),
\end{equation} 
where now, $\mu=\mu(\rhot^2,\bt^2)$, $\nu=\nu(\rhot^2,\bt^2)$. Note, by
the way, that $[\mu]=[\nu]=1$. The
coordinate transformation relating the line elements
(\ref{d_line_element}) and (\ref{nd_line_element}) is given by
\begin{eqnarray}
&& \bt^2=\ttilde^2-\left(\frac{\alpha^2\lambda}{\tau^2}\right) \zt^2, \\
&& \mbox{tanh}^2\left(\frac{\alpha\sqrt{\lambda}}
{\tau}\,\tilde\chi\right) =\left(\frac{\alpha^2\lambda}{\tau^2}\right) 
\frac{\zt^2}{\ttilde^2}.
\end{eqnarray}
In the rest of this work only dimensionless coordinates will be used,
therefore in order to simplify the notation we drop the tilde
$\;\widetilde{}\;\,$ from the coordinates $(\ttilde,\rhot,\varphi,\zt)$,
and $(\bt,\rhot,\varphi,\tilde\chi)$. 

In terms of the $(b,\rho,\varphi,\chi)$  coordinates the field equations 
take the form,
\begin{eqnarray}
&&\mu_{,\rho\rho}-\left(\frac{\alpha^2\lambda}{\tau^2}\right)\mu_{,bb}+
\rho^{-1}\mu_{,\rho}-\left(\frac{\alpha^2\lambda}{\tau^2}\right) 
b^{-1}\mu_{,b}=0, 
\label{field1}\\
&&\left(b^2-\left(\frac{\alpha^2\lambda}{\tau^2}\right)\rho^2\right)
\nu_{,\rho}=\frac{1}{2}\rho
b^2\left(\mu_{,\rho}^2+\left(\frac{\alpha^2\lambda}{\tau^2}\right)
\mu_{,b}^2\right)-\left(\frac{\alpha^2\lambda}{\tau^2}\right)\rho^2 b
\mu_{,\rho}\mu_{,b} \nonumber \\
&&\phantom{\left(b^2-\left(\frac{\alpha^2\lambda}{\tau^2}\right)\rho^2\right)
\nu_{,\rho}=XXXXX}-\left(\left(\frac{\alpha^2\lambda}{\tau^2}\right)\rho^2+
b^2\right)\mu_{,\rho}+2\left(\frac{\alpha^2\lambda}{\tau^2}\right)
\rho b \mu_{,b},
\label{field2} \\
&&\left(b^2-\left(\frac{\alpha^2\lambda}{\tau^2}\right)\rho^2\right)
\nu_{,b}=-\frac{1}{2}\rho^2
b\left(\mu_{,\rho}^2+\left(\frac{\alpha^2\lambda}{\tau^2}\right) 
\mu_{,b}^2\right)+\rho b^2
\mu_{,\rho}\mu_{,b} \nonumber \\
&&\phantom{\left(b^2-\left(\frac{\alpha^2\lambda}{\tau^2}\right)
\rho^2\right)\nu_{,\rho}=XXXXX}-\left(\left(\frac{\alpha^2\lambda}
{\tau^2}\right)\rho^2+b^2\right)\mu_{,b}+2\rho b \mu_{,\rho}.
\label{field3}
\end{eqnarray}
Equation (\ref{field1}) corresponds to the wave equation
(\ref{wave}),  and  is the integrability condition for (\ref{field2}) and
(\ref{field3}). These field equations suggest
considering boost-rotation symmetric solutions of the wave equation
(\ref{wave}) as seeds to construct boost-rotation symmetric
spacetimes. In fact, one can construct 
an infinite number of spacetimes of that kind which are regular everywhere
but blow up at infinity. These ``{seeds}'' lead to
spacetimes describing uniformly accelerated particles under the
action of an external field.  Bi\v{c}\'{a}k \& Schmidt
\cite{BicSch89a} have shown that the only boost-rotation symmetric
solution to (\ref{wave}) which decays to zero at null infinity is
$\mu=0$. Therefore, in order to construct (non-trivial) boost-rotation
symmetric spacetimes which are as asymptotically flat as possible, one
has to consider ``{seeds}'' $\mu$ which satisfy the wave equation with
sources moving along boost-rotation symmetric orbits, i.e.:
\begin{equation}
\label{wave_with_sources}
\Box \mu =8\pi \sigma,
\end{equation}
with
$\sigma=\sigma(\rho^2,b^2)=\sigma(\rho^2,t^2-\left({\alpha^2\lambda}
/{\tau^2}\right)
z^2)$, $[\sigma]=1$. This strategy will yield spacetimes which  will be singular at
least along the world lines of the uniformly
accelerated particles. Bi\v{c}\'{a}k and Schmidt \cite{BicSch89a})
have briefly discussed tachyonic boost-rotation symmetric sources.
Here, only sources moving with speed inferior to that of light
($t^2-\left({\alpha^2\lambda}/{\tau^2}\right) z^2<0$) will
be considered. Hence, the trajectories of the sources will remain
always below the roof. This fact will  acquire relevance later,  when
we try to identify the sources of the gravitational field of the
Newtonian limits.

Bi\v{c}\'{a}k \& Schmidt analysed the fall-off conditions that $\mu$ and
$\nu$ have to satisfy for the spacetime to have at least a local null
infinity. Moreover, they showed that for metric functions $\mu$ and $\nu$
satisfying the aforementioned asymptotic flatness conditions, it is possible
to add suitable constants to both $\mu$ and $\nu$ so that the resulting
spacetime has a global null infinity, in the sense that it admits smooth
spherical sections. Finally, they also showed that for a $\mu$ 
satisfying the fall off conditions and  vanishing
at the origin ($\mu(0,0)=0$), it is possible to construct a spacetime where
null infinity is regular except for four points: the ``good luck case''. These
are the points where the particles enter and leave the spacetime.  

As will be shown later ---and perhaps not so surprisingly---  asymptotic
flatness will appear to be a crucial ingredient for the existence of the
Newtonian limit of the spacetimes under consideration. Remarkably, 
non-asymptotically
flat solutions (i.e. those describing accelerated particles in uniform fields)
can be intuitively constructed from asymptotically flat solutions by sending
one of the particles to infinity and at the same time increasing its
corresponding mass parameter \cite{BicHoeSch83a}. 

The temporal metric and spatial metric of Ehlers' theory can be
constructed from the line element (\ref{d_line_element}) by performing
the  replacement $c\mapsto{1}/{\sqrt{\lambda}}$ everywhere in 
(\ref{d_line_element}), where $\lambda$ is the so-called causality
constant. Then, 
the required temporal $t_{ij}$ will
be obtained by multiplying the corresponding metric tensor $g_{ij}$ 
 by $\lambda$. 
In terms of the $(t,z,\rho,\chi$) variables, and up to the two first orders
in $\lambda$, these two
tensors read
\begin{equation}
t_{ij}=\left(\begin{matrix}
{\tau }^2\, e^{\nu } - \lambda{\alpha }^2\displaystyle{ 
z^2 }
    {t^{-2}}\left( e^{\mu } - e^{\nu } \right) & 0 & 0 &
\lambda \alpha ^2 z \,
   t ^{-1} \left( e^{\mu } - e^{\nu } \right) \cr 0 & 
- \lambda\alpha ^2e^{\nu }  & 0 & 0 \cr 
0 & 0 & -{
       {\alpha }^2\,\lambda \,{\rho }^2}{e^{-\mu }}   & 0 \cr 
\lambda \alpha ^2 z \,
   t ^{-1} \left( e^{\mu } - e^{\nu } \right) & 0 & 0 & 
-\lambda  {\alpha }^2 e^{\mu }  \cr 
\end{matrix}\right)
\end{equation}
\begin{equation}
s^ {ij}=\left(\begin{matrix}- {\lambda }\tau^{-2}{e^{-\nu }}  & 0 & 0 &\lambda\tau^{-2} \,z\,t^{-1} {\left( e^{-\mu } - e^{-\nu }
       \right) } \cr 
0 & \alpha ^{-2}e^{-\nu } & 0 & 0 \cr
 0 & 0 &{\alpha }^{-2}\,{{\rho }^{-2}} {e^
     {\mu }} & 0 \cr 
\lambda\tau^{-2} \,z\,t^{-1} {\left( e^{-\mu } - e^{-\nu }
       \right) } & 0 & 0 & \alpha^{-2}e^{-\mu } + 
   {\lambda\,{\tau }^{-2}\, z^2 }
{t^{-2}\left( e^{-\mu } - e^{-\nu } \right)} \cr
\end{matrix}
\right).
\end{equation}
It can be checked that $[t_{ij}]=T^2$, and
$[s^{ij}]=L^{-2}$. 

\section{The Newtonian limit of boost-rotation symmetric spacetimes}

The natural arena for the discussion of the Newtonian limit of
boost-rotation symmetric spacetimes is the region above the roof, i.e.
$t^2>\left({\alpha^2\lambda}/{\tau^2}\right) z^2$, where the spacetime is 
radiative. The reason for
this is that as one makes $\lambda \to 0$, the region below
the roof ($t^2<\left({\alpha^2\lambda}/{\tau^2}\right) z^2$) 
gets squeezed by the roof. That is, the
region below the roof disappears in the limit, while the roof becomes
the set $\{t=0\}$ ---see figure 2---.
\begin{figure}[t]
\label{figure:roof_limit}
\begin{center}
\includegraphics[width=.8\textwidth]{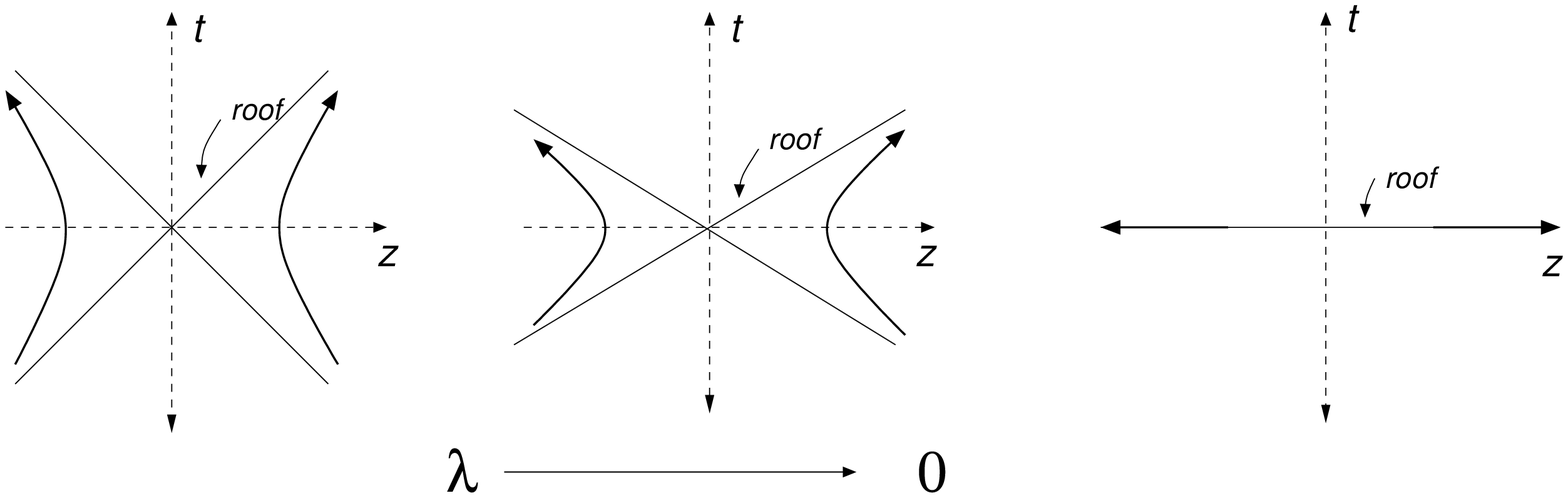}
\end{center}
\caption{The behaviour of the roof as the limit $\lambda\to 0$
  is taken.}
\end{figure}

\subsection{Necessary conditions for existence and consequences}

The conditions under which the hypothesis of theorem
\ref{existence:check} are satisfied for the boost-rotation symmetric
spacetimes described by either (\ref{d_line_element}) or
(\ref{nd_line_element}) are summarized in the following proposition.
\begin{proposition}
\label{lemma:existence}
\textbf{(Necessary conditions for the existence).} Consider the 
observer field $u^i=(\tau^{-1}e^{-\nu/2},0,0,0)$. 
Necessary conditions for
the existence of the Newtonian limit of the family
$(M(\lambda),g_{ij}(\lambda))$ of boost-rotation symmetric
spacetimes are
\begin{itemize}
\item[(i)] $\nu=\O(\lambda^0)$, $\mu=\O(\lambda^0)$, $\mu_{,\rho}=\O(\lambda)$,
$\mu_{,b}=\O(\lambda)$
\item[(ii)] $\nu_{,\rho}=\O(\lambda)$, $\nu_{,b}=\O(\lambda^0)$
\item[(iii)] $\nu_{,\rho\rho}=\O(\lambda)$, $\nu_{,\rho b}=\O(\lambda)$, 
$\nu_{,b b}=\O(\lambda^0)$.
\end{itemize}
\end{proposition} 

\textbf{Proof.} This follows from using theorem \ref{existence:check} and
direct inspection. Condition (i) arises from imposing hypothesis (a1)
and (a2) of theorem \ref{existence:check}. In particular, it is
important to note that
\begin{equation}
\frac{1}{\lambda}\det s^{ij}=-\frac{1}{\tau^2\alpha^6}
\left(\frac{1}{\rho^2}e^{-2\nu}\right).
\end{equation}
So,  the limit $\lambda \to 0$ may not exist if  
$\rho=0$. This peculiarity can be easily understood by recalling that the
boost-rotation symmetric spaces contain struts or conical
singularities on the axis $\rho=0$. These singularities represent the
particles undergoing uniform acceleration.

Along the same lines, condition (ii) of lemma \ref{lemma:existence}
arises from imposing hypothesis (b1) and (b2) of theorem
\ref{existence:check}. In particular, the condition
$\nu_{,\rho}=\O(\lambda)$ and $\nu_{,b}=\O(\lambda^0)$
appears from assuming that the limit
\begin{equation}
\lim_{\lambda\to 0} F=\lim_{\lambda\to 0}
\frac{1}{\lambda} \dd u^\bullet
\end{equation}
exists. This is because 
\begin{equation}
u^\bullet=t_{ij} u^i \dd x^j=\tau e^{\nu/2}\dd b=
\tau e^{\nu/2}\dd t+ \O(\lambda),
\end{equation}
where we have used the relation $b=t+\O(\lambda)$.
So, at the end of the day one has
\begin{equation}
\label{forces}
F=-\frac{\tau}{2\lambda}e^{\nu/2}\left(\nu_{,\rho} \dd t \wedge \dd\rho+
\nu_{,z} \dd t \wedge \dd z+\O(\lambda)
\right),
\end{equation}
and hence the need of $\nu_{,\rho}=\O(\lambda)$ and
$\nu_{,b}=\O(\lambda ^0)$ for the limit to
exist. In the remainder, we will show that this condition has
something to do with asymptotic flatness. Most boost-rotation symmetric 
spacetimes which are not at least locally asymptotically flat will not 
satisfy it.

Finally, condition (ii) stems from requiring the existence
of the limit as $\lambda\to 0$ of the Riemann tensor. \hfill $\Box$

Using the field equations (\ref{field1})-(\ref{field3}) it is not hard
to draw some immediate consequences of proposition
\ref{lemma:existence}.

\begin{proposition}
\label{lemma:consequences}
\textbf{(Consequences of the necessary conditions).} Assume that the
hypothesis of  
proposition \ref{lemma:existence} hold. Then,
\begin{itemize}
\item[(i)] 
\begin{equation}
\mu_{,\rho}=\O(\lambda), \qquad \mu_{,\rho\rho}=\O(\lambda);
\end{equation}
\item[(ii)] 
\begin{eqnarray}
&& \mu=\overset{\circ}{\mu}(b^2)+\lambda\Xi(\rho^2,b^2), \\
&& \nu =\overset{\circ}{\nu}(b^2)+\lambda \Phi(\rho^2,b^2),
\end{eqnarray}
where $\Xi=\O(\lambda^0)$, $\Phi=\O(\lambda^0)$, and according to the 
field equations 
$\overset{\circ}{\nu}_{,b}=-\overset{\circ}{\mu}_{,b}$;
\item[(iii)] 
\begin{equation}
\sigma=\lambda \Sigma(\rho^2,b^2),
\end{equation}
where $\Sigma=\O(\lambda^0)$. 
\end{itemize}
\end{proposition}

\textbf{Proof.} The $\lambda$-dependencies given in (i) come directly
from taking the limit $\lambda\to 0$ in the field equation
(\ref{field2}) and using the result in (\ref{field1}).
 Consequence (ii) follows from the last
discussion and from (ii) in proposition 
\ref{lemma:existence}. Finally, (iii) comes from using (i) and (ii) in
the wave equation with sources,
\begin{equation}
\mu_{,\rho\rho}+\rho^{-1}\mu_{,\rho}-\left(\frac{\alpha^2\lambda}{\tau^2}
\right)\mu_{,bb}-
\left(\frac{\alpha^2\lambda}{\tau^2}\right)b^{-1}\mu_{,b}=8\pi\sigma.
\end{equation} 
\hfill $\Box$

\subsection{The role of asymptotic flatness}

So far, asymptotic flatness of the boost-rotation
symmetric spacetimes (or the lack of it) 
has not entered our analysis. Ehlers argued that
the notion of asymptotic flatness should play a crucial
role in the existence of Newtonian limits \cite{Ehl98}. This
observation suggests the possibility that spacetimes describing
accelerated particles in uniform (gravitational) fields may not
possess a Newtonian limit proper. 

Following reference \cite{BicSch89a}, we will consider that a given
metric function $\mu$ ---satisfying a non-homogeneous wave equation--- is
said to be \emph{compatible with asymptotic flatness} if $\Omega^{-1}\mu$ is
smooth on null infinity, where $\Omega$ is a suitable conformal factor
defining null infinity. This means that it should have (at least) the
following asymptotic behaviour:
\begin{equation}
\mu=\mu_1\Omega+\O(\Omega^2).
\end{equation}
Consider for example the conformal factor,
\begin{equation}
\label{Omega}
\Omega=\frac{1}{\left({\tau^2}/{\alpha^2\lambda}\right)t^2-\rho^2-z^2}\,,
\end{equation}
which is valid for the region $\left({\tau^2}/{\alpha^2\lambda}\right)t^2>
\rho^2-z^2$. Now, consider a metric
function $\mu^*$ not compatible with asymptotic flatness. For example,
\begin{eqnarray}
&&\mu^*=\mu_{-1}\Omega^{-1}+\mu_0+\mu_1\Omega+\O(\Omega^2), \\
&&\phantom{\mu^*}=\mu_{-1}\left(\left(\frac{\tau^2}
{\alpha^2\lambda}\right)t^2-\rho^2-z^2\right) + \O(\Omega^0),
\end{eqnarray}
where $\mu_{-1}=\O(\lambda^0)$. Whence,
\begin{eqnarray}
\mu^*_{,\rho}=-2\rho\mu_{-1}+\cdots,
\end{eqnarray}
so that $\overset{\circ}{\mu}\,^*\!_{,\rho}\neq 0$, and accordingly
$\overset{\circ}{\nu}_{,\rho}\,^*\neq 0$. That is, condition
(ii) of proposition \ref{lemma:existence} is not satisfied, and thus
the boost-rotation symmetric spacetime to be obtained from the seed
function $\mu^*$ does not have a properly defined Newtonian limit.
This last result shows that, in fact, asymptotic flatness is a
prerequisite for the existence of a Newtonian limit. This should not be
a surprise because Newton's theory, as pointed out by Ehlers, is
actually a theory of isolated bodies. Several ways to impose it have
been suggested in the literature ---see in particular \cite{Tra66}---. 
Note, however, that a decay of the form
\begin{equation}
\mu=\mu_0+\mu_1\Omega+\O(\Omega^2),
\end{equation}
is still compatible with condition (ii) of proposition
\ref{lemma:existence}. This particular class of boost-rotation
symmetric spacetimes could still have a Newtonian limit. 

As should be expected, the asymptotic flatness of the general relativistic
solution leaves an imprint on the asymptotic behaviour of its Newtonian
limit. We have the following result.

\begin{proposition}
\label{proposition:af_consequences}
\textbf{(Consequences of asymptotic flatness).} Let $\mu$ be a solution of 
the wave equation with sources
(\ref{wave_with_sources}) compatible with asymptotic flatness. Then it
follows that
\begin{itemize}

\item[(i)] 
\begin{equation}
\overset{\circ}{\nu}_{,b}=\overset{\circ}{\mu}_{,b}=0;
\end{equation}

\item[(ii)] similarly, for fixed $t$ and $z$ one has, 
\begin{equation}
\Xi\sim \frac{C}{\rho^2}, \qquad \Phi\sim-\frac{C}{\rho^2},
\end{equation}
for large $\rho$, where $C$ is a constant.
\end{itemize}
\end{proposition}

\textbf{Proof.} Consider the conformal factor given in
(\ref{Omega}). It is not hard to see that
\begin{equation}
\Omega=\frac{1}{t^2}\left(\frac{\alpha^2\lambda}{\tau^2}\right) +
\frac{\rho^2+z^2}{t^4}\left(\frac{\alpha^2\lambda}{\tau^2}\right)^2+
O(\lambda^3).
\end{equation}
As mentioned before, a function $\mu$ compatible with asymptotic
flatness will be of the form
\begin{equation}
\mu=\mu_1\Omega=\O(\Omega^2),
\end{equation}
where $\mu_1=\O(\lambda^0)$, and $\partial_\Omega \mu_1=0$. Hence one
has
\begin{equation}
\label{mu:conformal}
\mu=\frac{\mu_1}{t^2}\left(\frac{\alpha^2\lambda}{\tau^2}\right)+O(\lambda^2).
\end{equation}
Accordingly
$\overset{\circ}{\mu}_{,b}=0$ ---cfr. (ii)
in proposition \ref{lemma:consequences}---. Thus, from
(ii) in (\ref{lemma:consequences}) one has
$\overset{\circ}{\nu}_{,b}=\overset{\circ}{\nu}_{,t}=0$. In
order to prove (iii), let us consider the conformal factor
\begin{equation}
\Omega'=\frac{1}{\rho^2+z^2-\left({\tau^2}/{\alpha^2\lambda}\right)t^2},
\end{equation}
in the domain in which it is positive. Then, asymptotically one has
$\Omega'\sim 1/\rho^2$, from which (ii) follows. \hfill $\Box$

\subsection{Interpretation and sources}

We are now in the position of discussing the Newtonian (gravitational)
potential of those boost-rotation symmetric spacetimes possessing a Newtonian
limit and describing the sources that give rise to them. Because of
proposition (\ref{lemma:consequences}) one has that the connection form ---see
equation (\ref{forces})--- is given by
\begin{equation}
F=- \frac{1}{2}{\Phi_{,\rho}} (e^{\nu /2} )\dd t\wedge
\dd\rho- \frac{1}{2}{\Phi_{,z}} (e^{\nu /2} )\dd t\wedge
\dd z+\O(\lambda), 
\end{equation}
and because of (i) in lemma \ref{lemma:existence} then $e^{\nu
  /2}=\O(\lambda^0)$ and therefore
\begin{equation}
\label{newtonian_potential}
\overset{\circ}{F}=-\frac{1}{2}\Phi_{,\rho}
\dd b\wedge \dd\rho-\frac{1}{2}\Phi_{,z}
\dd b\wedge \dd z
\end{equation}
The last result, along with theorem \ref{theorem:connection} and corollary
\ref{corollary:connection}, indicates that $-\Phi/2$ is the Newtonian
potential as measured by the rigid observers $u^i$. In
reference \cite{BicSch89a} it was argued that because of equation
(\ref{wave_with_sources}) and since ``$\Box\to \Delta$'' in the
 $\lambda\to 0$ limit, it is suggestive to interpret $\mu/2$ as 
a (Newtonian) gravitational potential. It was also noticed that in the
weak field approximation (i.e. weak sources) and for $t\approx 0$ one
has
\begin{equation}
g_{00}=e^\mu \approx 1+\mu,
\end{equation}
which brings further support to their point of view. Here we want to
make the point that this is only true if one enforces the requirement
of asymptotic flatness. To see this, divide field equation
(\ref{field2}) by $\tau^{-2}\lambda$ and take the limit as
$\lambda\to 0$.  Recalling that $b\to t$ as
$\lambda\to 0$ one obtains
\begin{equation}
\label{Phi_relates_to_Xi}
\tau^2t^2\Phi_{,\rho}=\frac{1}{2}\alpha^2\rho t^2(\overset{\circ}{\mu}_{,t})^2
-\tau^2t^2\Xi_{,\rho}+2\alpha^2\rho t \overset{\circ}{\mu}_{,t}.
\end{equation}
Thus, only if the boost-rotation symmetric spacetime is asymptotically
flat one will have that $\overset{\circ}{\mu}_{,t}=0$ and accordingly
$\Xi_{,\rho}=-\Phi_{,\rho}$. In addition to this,
$\Xi_{,z}=-\Phi_{,z}$ follows from an analogous discussion involving
equation (\ref{field3}).  For the Newtonian limit of the
non-asymptotically flat spacetimes described in \S 4.2, whose metric
seed function $\mu$ is of the form $\mu=\mu_0+\O(\Omega)$, the
interpretation of $\mu/2$ as a potential is not valid. This example
illustrates the subtle ---but crucial--- differences between the
notions of weak field approximation and Newtonian limit. The weak
limit approximation is still a curved spacetime theory valid for
sources which in some sense are moving slowly. Recall that the
boost-rotation symmetric spacetimes are time symmetric, so that the
uniformly accelerated particles start moving from rest at $t=0$.
Thus, in our case, this weak field approximation should only be
correct for $t\approx 0$, where they will still be moving slowly. On
the other hand, in the Newtonian theory there is no restriction to the
speed an object can attain, as long as it remains bounded for finite
times. Accordingly, it is valid for all times. Finally observe that if
one constructs a function $\mu$ such that
$\overset{\circ}{\mu}_{,b}=0$ then one would indeed have
$\Xi_{,\rho}=-\Phi_{,\rho}$ and $\Xi_{,z}=-\Phi_{,z}$.

Given a boost-rotation symmetric source
$\sigma=\sigma(\rho^2,t^2-\lambda{\alpha^2} z^2/{\tau^2})$, 
then by means of Kirchoff
integrals its is possible to construct retarded $(\mu_+)$ and advanced
$(\mu_-)$ fields of the form
\begin{equation}
\mu_\pm =\frac{1}{8\pi}\int_{\Real^4} \frac{\sigma\left(\rho^{\prime 2},
t^{\prime 2}-({\alpha^2\lambda}/{\tau^2}) z^{\prime
    2}\right)}{\sqrt{(\rho-\rho^{\prime})^2+(z-z^{\prime})^2}}\,\delta\left(
    \frac{\alpha}{\tau}\sqrt{\lambda}\sqrt{(\rho-\rho^{\prime})^2+
(z-z^{\prime})^2}\mp (t-t^\prime)\right)\dd^4x^\prime,
\end{equation}
where the integral is evaluated over the whole Minkowski spacetime. We
note that the boost-rotation symmetric source can be very complicated
and posses all sorts of multipole structures ---represented by
derivatives of the $\delta$-function---. The relevance
of these advanced and retarded solutions for our purposes 
is that by considering linear 
combinations of the functions $\mu_\pm$, one can
construct \cite{BicSch89a} a metric function $\mu$ that is an \emph{analytic}
function of $\rho^2$ and $t^2-\lambda z^2$ outside the sources, and
which is asymptotically regular at the fixed points of the
boost-symmetry on null infinity, namely
\begin{equation}
\mu= a\mu_+ +(1-a)\mu_-,
\end{equation}
where $a$ is suitable constant. Recalling that $\sigma=\lambda
\Sigma$, where $\Sigma=\O(\lambda^0)$ is analytic in $\lambda$ so that
$\lim_{\to 0} \Sigma(\rho^2,t^2-\lambda\alpha^2 z^2/\tau^2)$ does not depend
on $z$, one has that
\begin{eqnarray}
&&\lim_{\lambda\to 0} \frac{\mu}{\lambda}=\frac{a}{8\pi}\int_{\Real^4}
\frac{\overset{\circ}{\Sigma}\;\delta(t -t^\prime)}{\sqrt{(\rho-\rho^{\prime})^2+
(z-z^{\prime})^2}}\,
\dd^4x^\prime \nonumber \\ 
&&\phantom{\lim_{\lambda\to 0} \frac{\mu}{\lambda}XXXXXXXXXX}+
\frac{(1-a)}{8\pi}\int_{\Real^4}
\frac{\overset{\circ}{\Sigma}\;\delta(t -t^\prime)}{\sqrt{(\rho-\rho^{\prime})^2+
(z-z^{\prime})^2}} \,\dd^4x^\prime, \\
&&\phantom{\lim_{\lambda\to 0} \frac{\mu}{\lambda}}=\frac{1}{8\pi}
\int_{\Real^3}
\frac{\overset{\circ}{\Sigma}|_{t}}{\sqrt{(\rho-\rho^{\prime})^2+
(z-z^{\prime})^2}}\,
\dd^3x^\prime.
\end{eqnarray}
Thus, for an boost-rotation symmetric spacetime arising from an
asymptotically flat seed $\mu$ 
one gets that ---cfr. equation (\ref{Phi_relates_to_Xi})---,
\begin{equation}
\Phi=-\frac{1}{8\pi}\int_{\Real^3}
\frac{\Sigma(\rho^{\prime 2},t^2)}{\sqrt{(\rho-\rho^{\prime})^2+
(z-z^{\prime})^2}}
\dd^3x^\prime.
\end{equation}
Consequently, the potential $\Phi$ satisfies the Poisson equation with moving
sources:
\begin{equation}
\Delta (-\Phi/2) =4\pi \Sigma(\rho^2,t^2).
\end{equation}
Note, in particular, the minus sign in the source term. In the Newtonian
limit, the boost-rotation symmetric source has been replaced by a
moving cylindrically symmetric source analytic in $\rho^2$ and $t^2$. We
finish this discussion by recalling that the boost-rotation symmetric source
$\sigma$ is such that the amount of negative masses balances exactly the amount
of positive masses. That is,
\begin{equation}
\int_{\Real^3} \sigma\left(\rho^2,t^2-\left({\alpha^2\lambda}/{\tau^2}\right)
z^2\right) \dd^3 x'=0.
\end{equation}
It then follows  that $\Sigma(\rho^2,t^2)$ will inherit this property. Indeed,
\begin{equation}
\int_{\Real^3} \Sigma(\rho^2,t^2) \dd^3 x'=0.
\end{equation}

\section{Examples}

We now proceed to briefly discuss the Newtonian limits of some examples of
boost-rotation symmetric spacetimes. As noted in the introductory
sections, the first problem one has to face in
order to proceed with this discussion is how to transcribe the
exact general relativistic solutions ---which almost always are expressed in
terms of ``natural'' units for which $c=1$--- into the language of the theory
of frames. There is no canonical/unique way to do this. Indeed, one could for
example, introduce the causality constant $\lambda$ in such a way into the
Schwarzschild solution so that instead of the expected Newtonian limit of the
potential due to a point mass one obtains a vanishing potential. Thus, the
criteria for choosing a particular transcription rule is that it provides an
interesting Newtonian limit. In our particular case, consistent with
the discussion of section 3, we put forward the transition rules
\begin{eqnarray}
\label{transcription}
&& \rho_{BHS} \mapsto \alpha \rho, \\
&& z_{BHS} \mapsto \alpha z, \\
&& t_{BHS} \mapsto \frac{\tau}{\alpha \sqrt{\lambda}} t,
\end{eqnarray}
where $\rho_{BHS}$, $z_{BHS}$, $t_{BHS}$ are the
$\rho$, $z$, $\tau$ 
coordinates used by Bi\v{c}\'{a}k, Hoenselaers and Schmidt
\cite{BicHoeSch83a,BicHoeSch83b} (Bonnor \cite{Bon83} labels them
$\eta$, $\zeta$, $\tau$), and $\rho$, $z$ and $t$ are the dimensionless
coordinates used throughout most of this article. In order
to write the expressions in a more compact fashion, we define the
dimensionless quantity
\begin{equation}
\label{zeta}
\zeta=\left( \rho^2 +z^2 -\frac{\tau^2}{\alpha^2\lambda}t^2\right).
\end{equation}

\subsection{The two monopoles solution}
In \cite{BicSch89a}, Bi\v{c}\'{a}k \& Schmidt have constructed what is
arguably the 
simplest non-trivial example of a boost-rotation symmetric
``potential'' $\mu$: that obtained from the superposition of retarded and
advanced potentials due to two uniformly accelerated point particles, one with
a positive mass, and the other with a negative one. The function $\mu$ thus
obtained is by construction compatible with asymptotic flatness. Therefore,
the boost-rotation symmetric spacetime arising from it will also be 
asymptotically flat modulo the usual problems at the intersection of the light
cone through the origin with null infinity. Because of its asymptotic flatness,
the Newtonian limit is expected to exist, and therefore it can
be determined  by simply looking at $\mu$ (cfr. \S 4.2). In terms of the 
dimensionless
coordinates $(t,\rho,\chi,z)$ one has that
\begin{equation}
\label{dipole}
\mu= -\frac{4 m G \lambda}{\alpha \sqrt{(\zeta-1)^2+4\rho^2}},
\end{equation} 
where the constant $m$ has dimensions of mass and $\zeta$ given by
(\ref{zeta}). As explained previously, a non-trivial Newtonian limit
for boost-rotation spacetimes will only exist, strictly speaking, for
$t=0$ ---the sources of the wave equation (\ref{wave_with_sources}) and of
the associated limiting Poisson equation vanish ``above the
roof''---. Thus, the direct naive evaluation of the limit
$\lim_{\lambda \to 0} {\mu}/{\lambda}$ yields zero. In
order to extract a non-trivial limit out of (\ref{dipole}) we expand
around $t=0$, so that
\begin{equation}
\mu=-\frac{4 m G \lambda}{\alpha \sqrt{4\rho^2+(\rho^2+z^2-1)^2}}+
  \O(t^2).
\end{equation}
Thus, one could say that the Newtonian potential corresponding to the
relativistic two-monopole boost-rotation symmetric spacetime is, for
$t\approx 0$ given by
\begin{equation}
\label{dipole_newtonian}
\Phi\approx\frac{2 m G}{\alpha \sqrt{4\rho^2+(\rho^2+z^2-1)^2}}.
\end{equation}
\begin{figure}[t]
\label{figure:sources_1}
\begin{center}
\includegraphics[width=6cm,height=6cm]{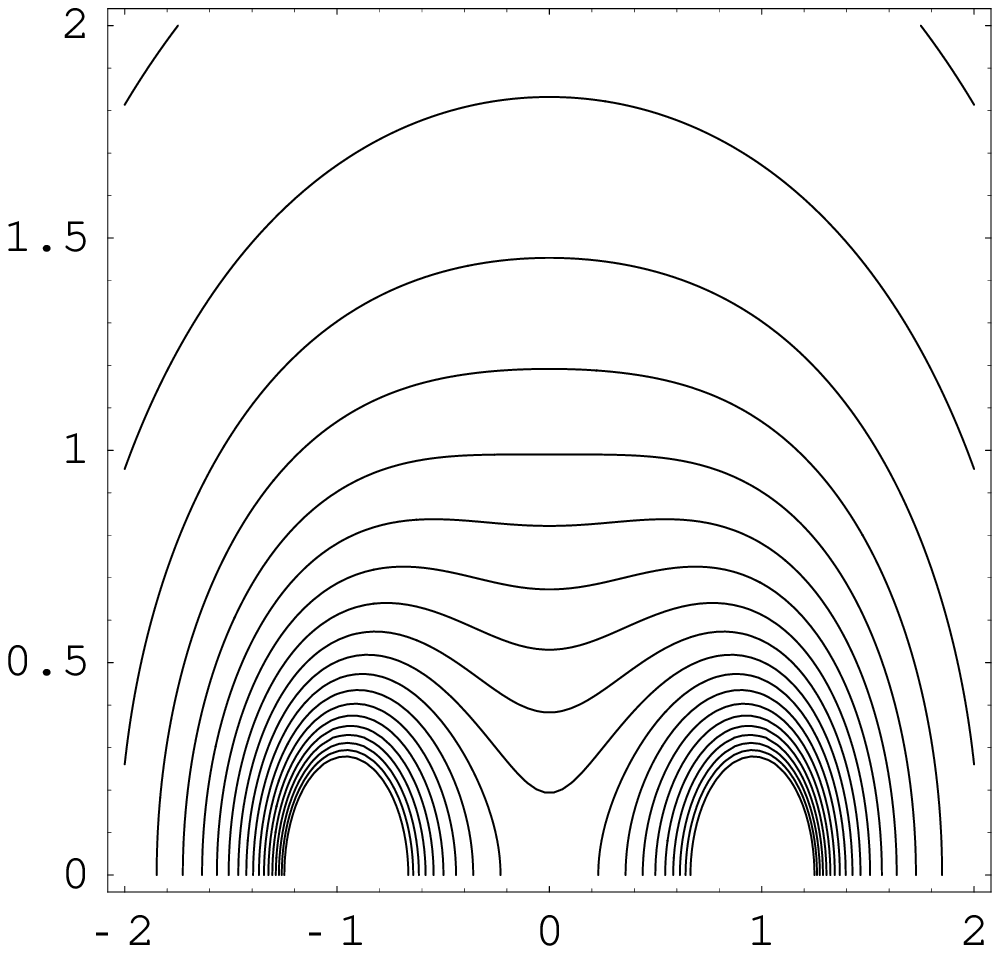}
\put(-78,-6){$z$}
\put(-180,88){$\rho$}
\end{center}
\caption{Level curves of the Newtonian potential due to the two
  monopole solution for $t\approx 0$.}
\end{figure}
The latter inherits the axial symmetry from the relativistic
solution. Furthermore, it is singular at two points lying on the $z$-axis,
$(\rho_{sing},\varphi_{sing},z_{sing})=(0,0,1)$ ---see figure 
3---. These
singularities are naturally identified with the presence of two
point-particles. Note that the potential is ---by construction--- time
independent. Thus, the Newtonian limit for early times is a strictly 
static Newtonian potential in which the sources are not
moving. Another remarkable feature of the solution is that, as can
readily be checked, the masses of the two point particles giving
rise to the Newtonian field have the same (positive) sign. 

\subsection{The Curzon-Chazy $(01)$-pole particles solution.}

This example of boost-rotation symmetric spacetimes was first given in
\cite{BicHoeSch83b}. It was constructed from the classical Bonnor-Swaminarayan
solution \cite{BonSwa64} by considering an appropriate limiting 
procedure. This solution is interpreted as the
superposition of a monopole particle and a dipole particle. 
Again, by construction, the
``seed'' function is compatible with asymptotic flatness so that a Newtonian
limit is bound to exist. In this case one has
\begin{equation}
\mu= -\frac{\scrd
  G\lambda}{\alpha^2}\frac{\zeta(\zeta-1)+2\rho^2}{\left((\zeta-1)^2+4\rho^2
\right)^{3/2}},
\end{equation}
where $\scrd$ a constant of dipolar nature ($[\scrd]=ML$), and $\zeta$ as in
(\ref{zeta}). Again, expanding $\mu$ around $t=0$ one finds that
\begin{equation}
\label{curzon_newton}
\Phi=\frac{\scrd G}{\alpha^2}\left(\frac{ (\rho^2+z^2)(\rho^2+z^2-1)+2\rho^2}
{(4\rho^2+(\rho^2z^2-1)^2)^{(3/2)}}\right).
\end{equation}
 The Newtonian potential is again singular at
$(\rho_{sing},\varphi_{sing},z_{sing})=(0,0,\pm
\sqrt{t^2+1})$. Looking at the level curves of the potential
(\ref{curzon_newton}) one can perceive the fingerprints of the ``dipole
structure'' of the point particles.  

\subsection{The Generalized Bonnor-Swaminarayan solution}
This solution was obtained in \cite{BicHoeSch83a} by using Ernst's
regularization procedure \cite{Ern78}. It describes two identical
particles symmetrically located with respect to the plane $z=0$ and
uniformly accelerated along the axis $\rho=0$. The interest of this
example for our purposes lies in the known fact that this spacetimes
is not asymptotically flat. In this case one has
\begin{equation}
\mu= \frac{\scrd G \lambda}{\alpha^2} \left(1-\frac{1}{
    \sqrt{(\zeta-1)^2+4\rho^2}}\right) + \left(
  2\rho^2-\zeta\right),
\end{equation}
where again $[\scrd]=ML$, and $\zeta$ given by (\ref{zeta}). It can be readily
seen that $\mu_{,\rho}=\O(\lambda^0)$, and thus, by propositions
\ref{lemma:existence} and \ref{lemma:consequences}, it does not possess a
proper Newtonian limit.
\begin{figure}[t]
\label{figure:sources_2}
\begin{center}
\includegraphics[width=6cm,height=6cm]{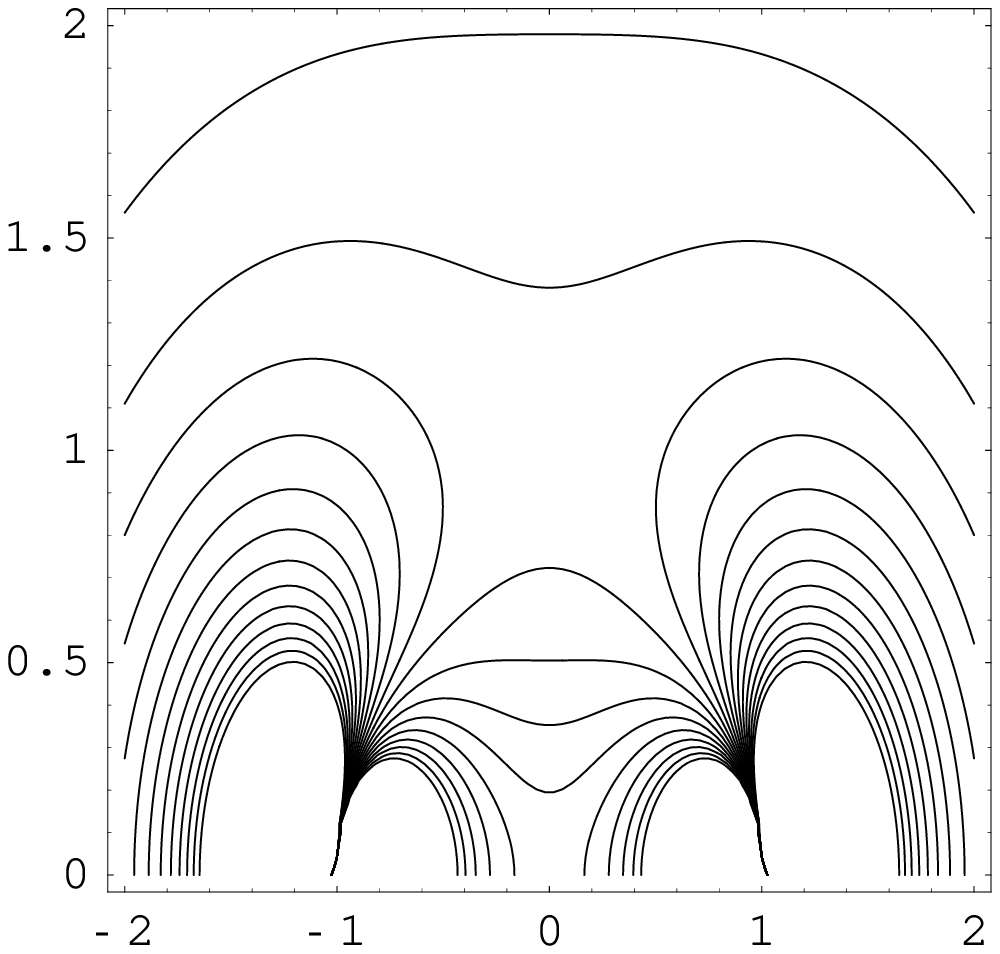}
\put(-78,-6){$z$}
\put(-180,88){$\rho$}
\end{center}
\caption{Level curves of the Newtonian potential due to the
  Curzon-Chazy $(01)$-pole solution.}
\end{figure}
\subsection{The C metric.}
We conclude our discussion of examples by considering the epitome of the
boost-rotation symmetric spacetimes: the C metric. Bonnor \cite{Bon83} was the
first  to cast it in the form which exhibits its boost-rotation symmetric
nature. The metric function $\mu$ arises from solutions to the wave equation
with sources which are uniformly accelerated rods moving along
the $z$ direction. In terms of our dimensionless coordinates one has
\begin{equation}
\label{c}
\mu= \ln \left( \frac{1}{4\omega^2} \frac{ \sqrt{ (\zeta-\beta)^2
      +4\beta\rho^2}+(\zeta-2\rho^2)-\beta}{\sqrt{ (\zeta-1)^2
      +4\rho^2}+(\zeta-2\rho^2)-1} \right).
\end{equation}
The classical parameters $m$ and $A$ are related to the parameters $\alpha$,
$\beta$ and $\omega$ via,
\begin{equation}
\alpha^2=2(z_2-z_1), \qquad \alpha^2\beta=2(z_2-z_3),
\end{equation}
where $z_1$, $z_2$, $z_3$ are solutions of
\begin{equation}
\label{cubic}
2A^4 z^3-A^2z^2+m^2=0. 
\end{equation}
It is assumed that the parameters $m$ and $A$ are such that the latter has 3
real solutions, and that $z_2$ is the biggest root. Finally,
\begin{equation}
\omega^2=4m^2A^{-6}\alpha^{-8}(\beta-1)^{-2}.
\end{equation}  

The points where $\mu$ is
singular can be identified with sources.  A simple calculation shows
then that this happens at $(0,0,z_{sing})$ with
$z_{sing}\in(-\sqrt{\beta},-1)\cup (1,\sqrt{\beta})$ if $\beta > 1$ or
$z_{sing}\in(-1,-\sqrt{\beta})\cup (\sqrt{\beta},1)$ if $\beta < 1$.
If $\beta=1$, then no singular points occur, and the C metric is in fact
Minkowski spacetime, that is, there are no sources. This seems to indicate
that the geometry is induced by the presence two rods of finite length
symmetrically located along the $z$-axis.  This agrees with the
description given by Bi\v{c}\'{a}k and Schmidt \cite{BicSch89a} on how
to construct the C-metric from a ``{s}eed''.

Now, a direct evaluation shows that the conditions
$\mu_{,\rho}=\O(\lambda)$ and $\mu_{,z}=\O(\lambda)$ needed for the
existence of a Newtonian limit of the C-metric do not hold unless
\begin{equation}
\label{beta}
\beta=1+\beta_1\lambda+\O(\lambda).
\end{equation} 
Assuming the latter one finds
\begin{equation}
\Phi = \frac{\beta_1}{2 \sqrt{4\rho^2+(\rho^2+z^2-1)^2}}, 
\end{equation}
at $t\approx 0$. That is, one recovers the potential of the two
monopoles solution.  This is in agreement with the results by Bonnor,
who concluded that in the weak field limit the C metric describes two
accelerated monopoles \cite{Bon83}.

\section{Conclusions}
We have discussed the Newtonian limit of boost-rotation symmetric
spacetimes. It has been shown that the existence, or not, of a Newtonian limit
depends on the asymptotic flatness of the relativistic spacetime. As discussed
in the main text, boost-rotation symmetric spacetimes posses two Killing
vectors: an axial one which is inherited by the Newtonian limit, and a boost
Killing vector. The boost Killing vector field is not inherited in any
clear way by the Newtonian limit as this symmetry is of relativistic nature.

In order to construct ``seed'' fields
$\mu$ which are analytical, one has to resort to suitable combinations of
advanced and retarded fields due to boost-rotation symmetric sources for which
the total amount of mass is zero. That is, one has to allow for the presence
of negative masses. The presence of regions of space containing negative
mass is preserved in the Newtonian limit in such a way that the total amounts
of positive and negative masses cancel exactly each other.

The standard interpretation of the boost-rotation symmetric spacetimes
regards them as models of uniformly accelerated particles. As has
been shown with the examples, the Newtonian limits are well defined for
all times, however the interpretation of particles moving in an
uniformly accelerated fashion is only valid for early times $t\approx
0$. The time dependence of Newtonian potential can be traced back
to the fact that the sources in the relativistic regime carry their own
source of motion: the struts or conical singularities joining
them. In other words, the motion of the  Newtonian sources is the Newtonian
consequence of the singularities in the relativistic boost-rotation symmetric
spacetimes. Summarizing, the Newtonian limits obtained exhibit several
unphysical features; however, all of them can be traced back to problems
already existing in the general relativistic solutions. Whether
the struts and strings appearing in the general relativistic solutions
here considered are ``terribly'' unphysical or not, is nevertheless a
matter of taste.

Finally, in \S 4.3 it has been shown that the potential suggested by a weak
field analysis coincides, in the case where asymptotic flatness is required,
with the Newtonian potential obtained through our analysis. There has been
some discussion in the literature ---cfr. \cite{BicSch89a,Bon83}--- in what
regards writing the potentials as a combination of advanced and retarded
fields. This discussion lies beyond the realm of our analysis, because in a purely
Newtonian theory  information travels with infinite speed. In order to look
at these effects one would have to look at the post-Newtonian expansions of
the solutions.

\section*{Acknowledgements}
We would like to thank Profs. J.M.M. Aguirregabiria, W.B. Bonnor, L.
Bel, J. Bi\v{c}\'{a}k, Dr. B. Coll and Profs. J.B. Griffiths, J.M.M.
Senovilla and B. Schmidt for their interest in this work and for
enrichening discussions and remarks in previous versions of this
article. J.A.V.K. wishes to thank the Dept. of Theoretical Physics of
the University of the Basque Country for their hospitality during the
completion of part of this work. R.  L.'s work is supported by the
Spanish Ministry of Science and Technology jointly with FEDER funds
through research grant BFM2001-0988, the University of the Basque
Country through research grant UPV172.310G02/99, and the Basque
Government through fellowship BFI01.412. J. A. V. K. is currently a Lise
Meitner fellow of the Austrian FWF (M690-N09).

\section*{Appendix: the axioms of Ehlers' theory}

The axioms of Ehlers' frame theory and their consequences are
discussed extensively for example in \cite{Lot88}. However, this
reference is not so readily available. Therefore, for the sake of completeness we
present them here. The axioms shown here are an adapted version of those
appearing in Lottermoser's thesis. The first set of axioms deals with the
objects that the frame theory will attempt to describe.

\begin{axiom}
\textbf{(On the objects described by the frame theory).} The mathematical
  objects of the theory are collections
  $(M,t_{ij},s^{ij},\Gamma^i_{jk},
T^{ij},\lambda,G)$
 for which the following holds:
\begin{itemize}
\item[i)] $M$ is a 4-dimensional Hausdorff manifold endowed with a
  connection,
\item[ii)] $t_{ij}$, $s^{ij}$ and 
$T^{ij}$ are symmetric
  tensor fields on $M$ (\textbf{temporal metric, spatial metric, 
matter tensor}),
\item[iii)] $\Gamma^i_{jk}$ is a torsion-free linear 
connection on $M$
  (\textbf{gravitational field}),
\item[iv)] $\lambda$ and $G$ are real numbers (\textbf{causality and
    gravitational constant}).
\end{itemize}
\end{axiom}

The axioms require some minimal differentiability conditions. Thus,
$t_{ij}$, $s^{ij}$, $T^{ij}$ and
$\Gamma^i_{jk}$ should at least be $C^1$ and $M$ at least a $C^3$
manifold. 

The next set of axioms describes, by means of observable quantities, how the objects of the frame theory ought to
be interpreted.

\begin{axiom}
\textbf{(Physical interpretation)}
\begin{itemize}
\item[i)] \textbf{Observers} move along timelike curves, that is, curves
  such that their tangent vector $u^i$ satisfies 
$t_{ij}u^i u^j>0$ everywhere. The \textbf{space directions} of an 
observer are the tangent vectors $v^i\neq 0$ along the curves which are 
orthogonal to $u^i$ with respect to $t_{ij}$: 
$t_{ij}u^i v^j=0$. 

\item[ii)] \textbf{Time intervals} are defined along timelike curves. Let $s$
  be the curve parameter. Then for the infinitesimal time interval 
the following holds:
\begin{equation}
dt=\sqrt{t_{ij}u^i u^j} ds.
\end{equation}
\item[iii)] \textbf{Space intervals} are defined along spacelike curves,
  i.e. curves such that for their  tangent vector $v^i$ there
  exists a 1-form $\omega_i$ which satisfies
\begin{equation}
v^i=\omega^i_\bullet,
\end{equation}
and $s^{ij}\omega_i\omega_j>0$. If $s$ is again the curve
parameter, then the \textbf{spatial interval} is given by
\begin{equation}
dl=\sqrt{s^{ij}\omega_i\omega_j}ds.
\end{equation}
\end{itemize}
\end{axiom}

The following 2 axioms establish the conditions a mathematical model
should satisfy in order to be called a ``solution of the frame
theory''.

\begin{axiom} 
\textbf{(Metric axioms)}
\begin{itemize}
\item[i)] At each point of the spacetime, there exists at least one timelike
  vector,
\begin{equation}
\forall p \in M \;\;\exists u^i\in T_p M:\;\; 
t_{ij} u^i u^j>0.
\end{equation}

\item[ii)] At each point of the spacetime, the temporal metric is positive
  definite at each observer orthogonal subspace of the
  cotangent space. That is, 
\begin{equation}
\forall p \in M, \;\;u^i \in T_p M,\;\; 
t_{ij}u^i u^j >0
\end{equation}
$s^{ij}$ is positive definite in the set $H^+(u^i)=\{\omega \in
T_p^* M|\;\; \omega_i v^j =0 \}$.

\item[iii)] the temporal and spatial metrics are related to each other via,
\begin{equation}
t_{ij}s^{jk}=-\lambda \delta_i^k.
\end{equation} 
\end{itemize}
\end{axiom}

\begin{axiom}
\textbf{(Connection axioms)}
\begin{itemize}
\item[i)] The connection $\Gamma^i_{jk}$ is compatible with the
  temporal and spatial metrics
  metrics, i.e.
\begin{equation}
t_{ij;k}=0, \qquad 
s^{ij}_{\phantom{ij};k}=0
\end{equation}
\item[ii)] the curvature tensor
  $R^i_{jkl}=
2(\Gamma^i_{j[l,k]}+
\Gamma^i_{m[k}\Gamma^m_{l]j})$ 
of $\Gamma^i_{jk}$ possesses the property,
\begin{equation}
R^{i\phantom{j}k}_{\phantom{i}j\bullet l}=
R^{k\phantom{l}i}_{\phantom{k}l\bullet j}.
\end{equation}
\end{itemize}
\end{axiom}


\end{document}